\documentclass[a4paper, 10pt, conference]{IEEEtran}
\IEEEoverridecommandlockouts
\usepackage{cite}
\usepackage{amsmath,amssymb,amsfonts}
\usepackage{algorithmic}
\usepackage{graphicx}
\usepackage{textcomp}
\usepackage{xcolor}
\usepackage{booktabs}
\usepackage{comment}
\usepackage{threeparttable}
\usepackage{subfig}

\usepackage{tikz}
\usetikzlibrary{positioning}

\def\BibTeX{{\rm B\kern-.05em{\sc i\kern-.025em b}\kern-.08em
    T\kern-.1667em\lower.7ex\hbox{E}\kern-.125emX}}
\begin{document}

\title{The Impact of Spatial Misalignment and Time Delay on Collaborative Presence in Augmented Reality\\
}
%The Impact of Spatial Misalignment and Time Delay on Collaborative Tasks in Augmented Reality

\makeatletter
\newcommand{\linebreakand}{%
  \end{@IEEEauthorhalign}
  \hfill\mbox{}\par
  \mbox{}\hfill\begin{@IEEEauthorhalign}
}
\makeatother

\author{
\IEEEauthorblockN{Michael Stern$^{1}$, Maurizio Vergari$^{2}$, Julia Schorlemmer$^{1}$, Francesco Vona$^{1}$, \\David Grieshammer$^{1}$, Jan-Niklas Voigt-Antons$^{1}$}
\IEEEauthorblockA{$^1$Immersive Reality Lab, Hamm-Lippstadt University of Applied Sciences, Hamm, Germany\\
$^2$Quality and Usability Lab, Technische Universität Berlin, Berlin,  Germany}
}

\maketitle

\newcommand\copyrighttext{%
\footnotesize \textcopyright 2025 IEEE. Personal use of this material is permitted. Permission from IEEE must be obtained for all other uses, in any current or future media, including reprinting/republishing this material for advertising or promotional purposes, creating new collective works, for resale or redistribution to servers or lists, or reuse of any copyrighted component of this work in other works. M. Stern, M. Vergari, J. Schorlemmer, F. Vona, D. Grieshammer and J. -N. Voigt-Antons, "The Impact of Spatial Misalignment and Time Delay on Collaborative Presence in Augmented Reality," 2025 17th International Conference on Quality of Multimedia Experience (QoMEX), Madrid, Spain, 2025, pp. 1-7, doi: 10.1109/QoMEX65720.2025.11219982.}

\newcommand\copyrightnotice{%
\begin{tikzpicture}[remember picture,overlay,shift={(current page.south)}]
\node[anchor=south,yshift=10pt] at (0,0) {\fbox{\parbox{\dimexpr\textwidth-\fboxsep-\fboxrule\relax}{\copyrighttext}}};
\end{tikzpicture}%
}
\copyrightnotice

\begin{abstract}
Precise temporal and spatial alignment is critical in collaborative Augmented Reality (AR) where users rely on shared visual information to coordinate actions. System latency and object misalignment can disrupt communication, reduce task efficiency, and negatively impact the overall user experience. While previous research has primarily focused on individual AR interactions, the impact of these inconsistencies on collaboration remains underexplored. This article investigates how user experience and task load are affected by object misalignment and time delay in a shared AR space. To examine these factors, we conducted an experiment with 32 participants, organized into 16 pairs, who collaboratively completed a spatial placement task. Within each condition, both participants alternated roles, taking turns as the leader—providing verbal placement instructions—and the builder—executing the placement. Six conditions were tested, manipulating object alignment (perfectly aligned vs. randomly misaligned) and time delay (0s, 0.1s, 0.4s). The misalignment was applied randomly to each virtual object with a shift of ±20 cm on every axis to create a clear distinction in spatial perception. User experience and task load were assessed to evaluate how these factors influence collaboration and interaction in AR environments. Results showed that spatial misalignment significantly increased perceived workload (NASA-TLX) and lowered user ratings in Pragmatic quality and Attractiveness (UEQ), while time delay had a more limited effect. These findings highlight the critical role of spatial accuracy in maintaining collaboration quality in AR. 
\end{abstract}


\section{Introduction}
Augmented Reality (AR) has rapidly emerged as a transformative technology for enhancing collaborative tasks by integrating digital content with the physical world. In shared AR environments, precise spatial alignment and minimal system latency are essential for ensuring that virtual elements seamlessly integrate with real-world settings, thereby fostering effective communication and coordinated action among users. However, even small deviations in object alignment or delays in system response can disrupt the shared experience, leading to increased task load and diminished user satisfaction.

Recent research has explored the role of AR in diverse fields, ranging from education and product design to industrial applications, highlighting its potential to improve collaborative interactions \cite{Yu2022Duplicated,Sereno2020Collaborative, Marques2021A}. Despite these advances, the combined effects of spatial misalignment and time delay on collaborative performance remain underexplored. While prior studies have separately addressed issues related to object misalignment and system latency, a comprehensive understanding of how these factors interact to influence user experience and task performance is lacking.

This study addresses this gap by investigating the impact of both spatial misalignment and time delay on collaborative tasks in AR. Specifically, we examine how these factors affect users' perceived workload and overall experience during a Collaborative Tower Building Task. By manipulating the alignment of virtual objects and introducing controlled time delays, we aim to elucidate the challenges faced by users when collaborating in imperfect AR environments. The results of this study have the potential to inform the design of more robust AR systems that can better accommodate the inherent imperfections of current technology, ultimately improving collaborative performance.

In the following sections, we describe the experimental design, task protocol, and procedures used to assess the effects of these variables on user experience and task load.

\section{Related Work}
\label{sec: related_work}
Augmented Reality (AR) systems offer unique opportunities to enhance collaboration by allowing multiple users to interact with shared 3D virtual objects in real-time. These interactions can enrich face-to-face communication or create a sense of virtual co-location for remote collaborators \cite{Yu2022Duplicated,Sereno2020Collaborative}. The seamless integration of virtual and real-world elements establishes common ground among users, fostering shared understanding and improved collaborative performance in various tasks \cite{Marques2021A}. AR has already demonstrated potential across several domains, including education, product design, and industry. In higher education settings, AR-enabled collaborative learning has been shown to enhance educational outcomes and positively influence collaboration experiences \cite{Costa2022Augmented}. Within industrial contexts, AR significantly supports human-robot collaboration, improving both task efficiency and safety measures \cite{Upadhyay2024Collaborative}. Similarly, in product design, AR allows multidisciplinary teams to collaboratively engage in concurrent design activities, enabling real-time modifications and interactive prototyping \cite{Poretski2021Physicality}.

However, despite these opportunities, collaborative AR experiences are often hindered by spatial misalignment—one of the most critical challenges faced by AR systems. Misalignment can reduce the ability to accurately perceive the location and activities of remote collaborators, undermining the shared spatial understanding crucial for effective teamwork \cite{Fink2022Re-locations}. It also interferes with path integration, limiting users’ capacity to construct coherent global spatial representations and navigate effectively within the virtual environment \cite{Lei2023Visual}. Moreover, misaligned perspectives often require additional mental transformations, leading to increased cognitive load and reduced task efficiency \cite{Pouliquen-Lardy2016Remote}. These viewpoint realignment efforts can introduce response conflicts, further complicating spatial judgments and decisions \cite{Sohn2003Viewpoint}. Another important concern is depth perception distortion, which arises when misalignment between eye position and AR overlays disturbs users’ ability to judge spatial relationships accurately \cite{Tong2022The}. Collectively, these issues hinder communication efficiency, as collaborators must invest extra effort to reconcile differing spatial perspectives, potentially slowing down interactions and causing misunderstandings \cite{Galati2013The}.

Previous research examined the effects of virtual element misalignment in collaborative augmented reality (AR) environments, focusing on how differing perceptions of virtual objects, positional synchrony, and avatars influence collaboration \cite{Vona2024Misalignment}. The findings indicated that while divergent perceptions did not significantly disrupt communication, positional synchrony was crucial for improving collaboration quality. Conversely, avatars had a limited impact, suggesting they cannot effectively compensate for misalignment issues.

A variety of measures can be used to assess the quality of systems as experienced by users \cite{arndt2017exploring}. In addition to traditional subjective measurement methods, brain signal measurement has been effectively utilized to estimate user satisfaction \cite{antons2015neural}.

In addition to spatial challenges, latency represents a critical factor influencing collaborative dynamics in AR. High latency can degrade task performance and mutual understanding, particularly in tasks involving real-time visual and verbal communication \cite{Becher2019Negative}. Moreover, maintaining synchronization and consistency of user states is essential for coherent social interactions. Techniques such as mobile edge computing and group motion prediction have been proposed to reduce latency and ensure temporal consistency across users, thus enhancing collaborative experiences in Social Extended Reality (XR) settings \cite{Park2018Minimizing, Hsiao2022Latency}. Furthermore, latency plays a pivotal role in determining the Quality of Experience (QoE) in immersive environments. While its effects are well-studied in traditional communication systems, its impact on Social XR is still an emerging area of inquiry. New frameworks that map latency sources to perception and acceptance thresholds are instrumental in predicting and improving QoE \cite{Cortés2024Understanding}.

Both spatial misalignment and latency issues inherently increase cognitive load and distraction, potentially affecting not only collaborative effectiveness but also user safety, especially in public and dynamic environments. Recent research highlights how design choices in AR interfaces—particularly the quantity and visualization characteristics of virtual markers—can significantly affect users’ sense of safety by altering cognitive workload and situational awareness \cite{Vergari2023Safety}. Specifically, excessive visual markers or points of interest (POI) can clutter the AR interface, leading to increased distraction and reduced ability to perceive and respond to real-world hazards, such as pedestrians, vehicles, or obstacles. Although adjusting visual parameters such as transparency, size, and color did not yield statistically significant results, users indicated that these visual adjustments influenced their perceived safety. Thus, optimizing AR interface design to reduce distraction and cognitive overload becomes crucial, directly connecting usability and safety considerations with spatial and latency challenges identified previously \cite{Vergari2023Safety}. Furthermore, the social environment in which AR experiences occur, as well as the focus of user interactions, significantly impacts both user experience and social acceptability. Recent findings indicate that AR applications used in crowded public settings are perceived as less socially acceptable compared to more intimate, uncrowded settings \cite{Cocchia2024Social}.

A related body of work has explored the relationship between presence and reaction time in mixed reality (MR) environments. Research shows that users with a higher sense of presence typically exhibit faster reaction times to stimuli, suggesting that reaction time could serve as an objective indicator of presence. Consequently, any factors—such as misalignment or delay—that impair reaction time may also diminish users’ felt sense of presence \cite{Chandio2024Human,Chandio2023Investigating}.

%Collaboration in AR/VR
%Interaction?
%Spatial Misalignment in AR/VR
%Time Delay in AR/VR
%User Experience and Task Load?

\section{Methods}
\subsection{Study Design}

%%%%%

This study follows a 2-factor within-subject experimental design with randomization of conditions. Time delay and spatial misalignment are the independent variables, while user experience and task load are the dependent variables. The time delay variable has three levels (0.0s, 0.1s, 0.4s), while the spatial misalignment variable has two levels (perfectly aligned vs. randomly misaligned), resulting in six conditions as shown in Table~\ref{conditions}. 

The three values for the time delay were chosen based on previous work that explored interaction delays in the range of 0 to 0.8s \cite{brunnstrom2019quality, allison2001tolerance,kelkkanen2023hand}. Meanwhile, pilot tests established misalignment values, which brought the random shift of each block within ±20 cm on each axis to create a clear distinction in spatial perception, while still ensuring a plausible experience. 

Based on the related work discussed (see Sec. \ref{sec: related_work}), we formulated the following research questions: 
\begin{itemize} 
    \item RQ1: How does spatial misalignment affect perceived task load and user experience in collaborative AR tasks? 
    \item RQ2: How does time delay affect perceived task load and user experience in collaborative AR tasks? 
    \item RQ3: What are the combined effects of spatial misalignment and time delay on task load and user experience in collaborative AR tasks?
\end{itemize}

\begin{table}[h!]
\centering
\caption{Conditions A-F: Virtual Object Alignment and Specified Time Delay}
\label{conditions}
\begin{tabular}{lccc}
\toprule
Condition  & Alignment           & Time Delay \\
\midrule
A              & Perfectly Aligned   & 0.0        \\
B              & Perfectly Aligned   & 0.1        \\
C              & Perfectly Aligned   & 0.4        \\
D              & Misaligned          & 0.0        \\
E              & Misaligned          & 0.1        \\
F              & Misaligned          & 0.4        \\
\bottomrule
\end{tabular}
\end{table}

\begin{figure*}[h]
    \centering
    \includegraphics[width=0.66\columnwidth]{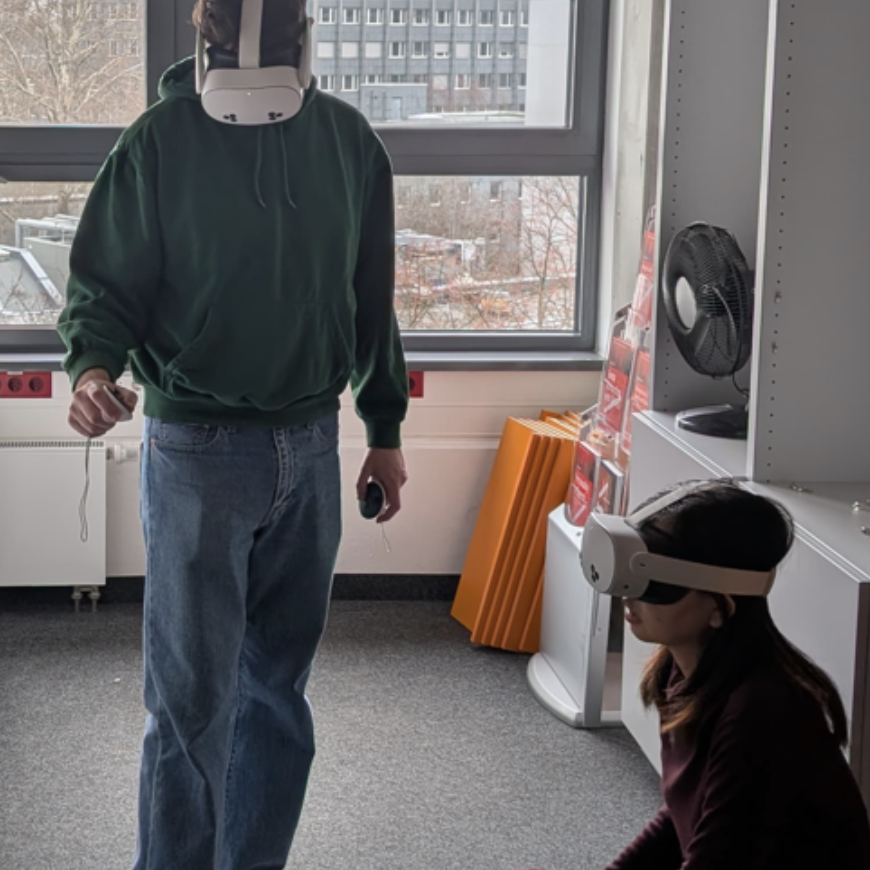}
    \includegraphics[width=0.66\columnwidth]{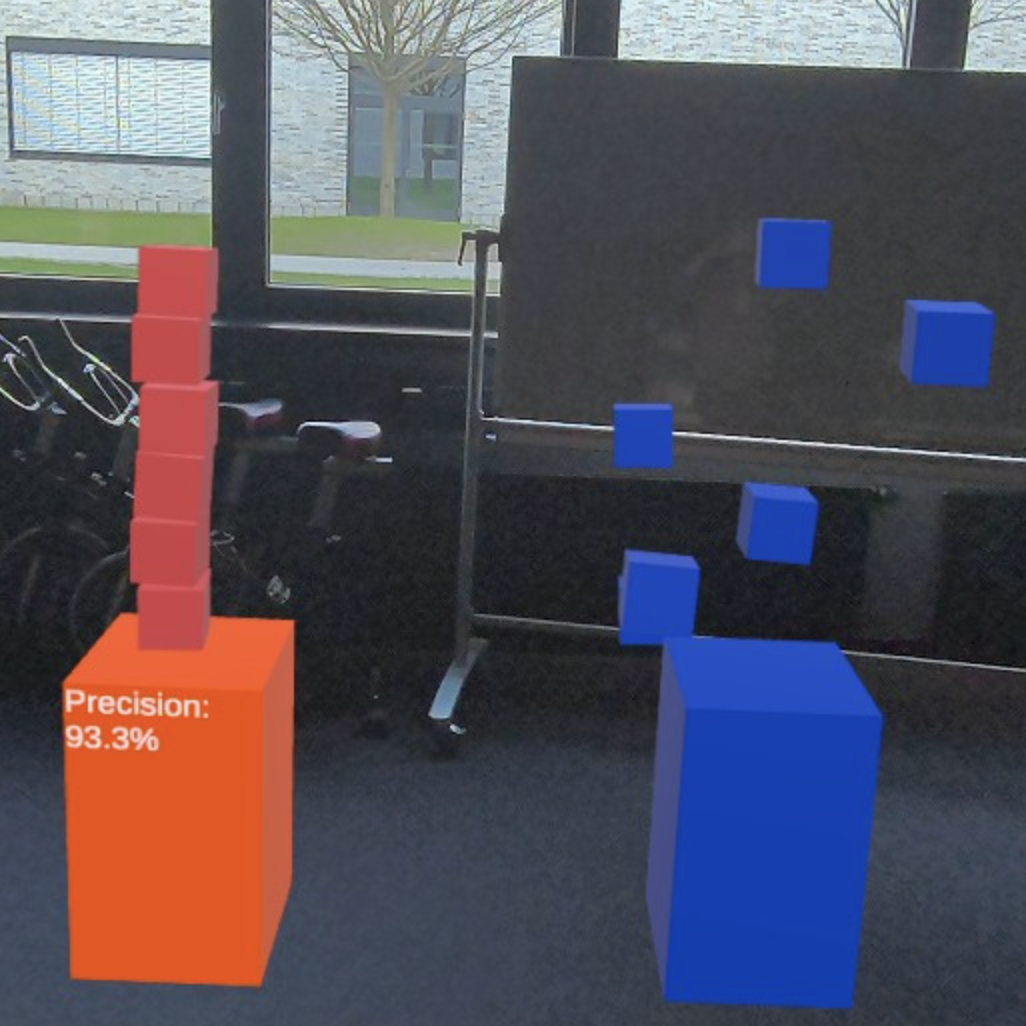}
    \includegraphics[width=0.66\columnwidth]{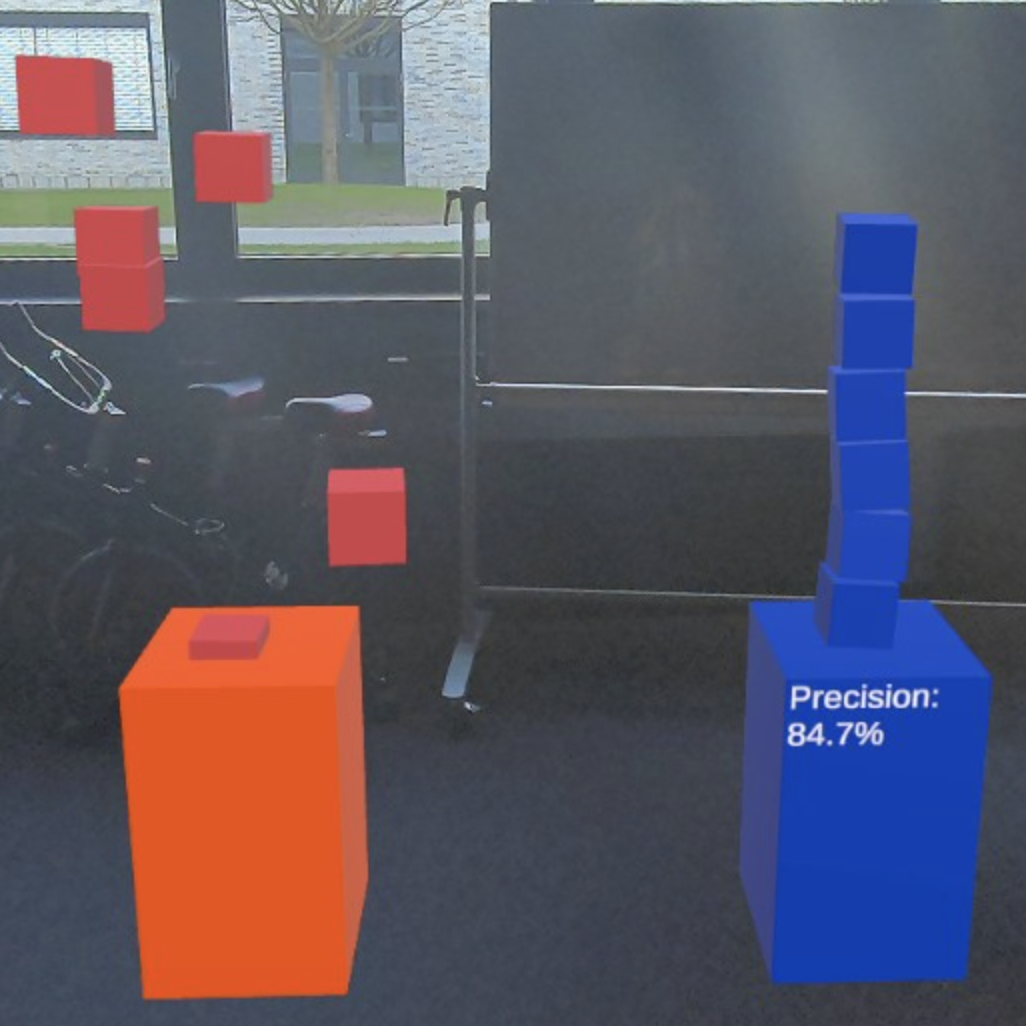}
    
    \caption{Builder and Leader collaborating on building the red tower (left); Leader's view (center); Builder's view (right).}
    \label{fig:task}
\end{figure*}

\subsection{Apparatus}
The VR application was implemented in Unity 3D and ran on two Meta Quest 3s. The application utilized Oculus Integration for interaction with virtual objects and employed custom scripts to introduce time delays and spatial misalignment during the task. Multiplayer functions were enabled via NORMCORE’s multiplayer plugin\footnote{https://normcore.io/}. Questionnaires were administered on a laptop using a self-developed tool, while the consent form and introduction sheet, detailing the experiment’s aim, procedure, and modalities were provided on paper.

\subsection{Collaborative Tower Building Task}
The Collaborative Tower Building Task is an activity that requires cooperative work between two people who assume the roles of builder and leader in turns while each builds a tower by stacking blocks. Each participant has to build a blue or red tower. The leader and the builder see the same set of blocks but have varying perceptions of spatial misalignment and time delay. 

In misaligned conditions, the builder sees their blocks with small positional offsets within a specified range. The leader sees the blocks and the building tower in their corresponding positions. The role of the leader is to guide the builder by giving verbal instructions so that the blocks are positioned correctly. In misaligned conditions, the builder should consider the leader's instructions to build a tower despite the misalignment successfully. 

The time delay condition further complicates the task in that the observation by the leader of the builder's action is delayed by a fixed period (e.g., 0.0, 0.1, or 0.4 seconds), making precise adjustments more difficult to achieve. 

To add a competitive layer to the task, both players can see each other's precision score as a percentage, but not their own. This design choice encouraged teamwork and communication between the two people. The collaboration of the leader and the builder for the red tower can be found in Fig. \ref{fig:task}-left. An example of the leader's view of the red tower in a misaligned condition can be found in Fig \ref{fig:task}-center. While, the red blocks misalignment in the Builder's view can be found in Fig. \ref{fig:task}-right.

\subsection{Participants}
Participants were recruited from university students. The final sample comprised 32 individuals, with an average age of 31.03 years (\textit{SD} = 7.55). The sample included 65.6\% female participants (n = 21) and 34.4\% male participants (n = 11). In terms of prior experience with related technologies or tasks, 21.9\% (n = 7) reported no experience, 53.1\% (n = 17) had limited experience, and 25.0\% (n = 8) reported moderate to extensive experience. Participants’ affinity for technology interaction (ATI) was assessed using the ATI scale \cite{Franke2019ATI}. The sample showed an overall mean score of 4.08 (\textit{SD} = 0.97), reflecting a moderate to high level of comfort and familiarity with technology.

\subsection{Procedure}
On average, the experiment took between 60 and 90 minutes. The experiment was conducted in a lab setting, where each pair of participants was invited at a different time slot. Upon arrival, the moderator welcomed the participants, provided a written introduction to the study, and obtained their consent and compensation-related information. Prior to the main experiment, participants completed a laptop pre-questionnaire that gathered demographic data, details on previous extended reality experience, and included the ATI scale \cite{Franke2019ATI}.

Following the pre-questionnaire, participants were assigned a condition and given a soft deadline of 8 minutes to complete it. The experiment comprised 6 conditions presented in a randomized order. In each condition, participants engaged in the Collaborative Tower Building Task, alternating roles as builders and leaders—working both on constructing their own tower and guiding their partner to build theirs. After completing each condition, participants filled out questionnaires to assess various aspects of their experience, including the User Experience Questionnaire (UEQ) \cite{Schrepp2017UEQ} and the NASA Task Load Index (NASA-TLX) \cite{Hart1988NASA}. The experiment concluded once all six conditions and the corresponding questionnaires had been completed.

\begin{comment}
\begin{figure}[h!]
    \centering
    \includegraphics[width=\columnwidth]{figures/participants_cropped.png}
    \caption{Collaborating participants during the experiment.}
    \label{fig:participant1}
\end{figure}
\end{comment}

\section{Results}
A series of statistical analyses were conducted to examine how time delay and object misalignment affected participants’ perceived task load and user experience during collaborative AR interactions. Descriptive statistics were first computed for all six experimental conditions to provide an overview of the responses across the NASA-TLX and UEQ scales. To assess potential differences between conditions, repeated-measures analyses of variance (RM-ANOVAs) were then performed, followed by post-hoc comparisons where appropriate. These analyses aimed to explore whether the manipulated factors influenced participants’ task demands and their subjective experience within the shared AR environment.

\subsection{Descriptive statistics}
Descriptive statistics were calculated for the NASA Task Load Index (NASA-TLX) and the User Experience Questionnaire to provide an overview of participants’ perceived workload and user experience across the six experimental conditions. This analysis serves to identify general trends and differences in subjective evaluations prior to conducting inferential analyses.

\subsubsection{NASA Task Load Index (NASA-TLX)}

Figure~\ref{fig:tlx_total_scores} illustrates the mean NASA-TLX total workload scores across the six experimental conditions, each combining different levels of spatial alignment and time delay. As shown in the figure, conditions with perfect alignment (A–C) yielded lower workload ratings overall compared to the misaligned conditions (D–F). The lowest workload was observed in Condition A (perfect alignment, no delay; \textit{M} = 39.19, \textit{SD} = 19.51), while the highest workload was reported in Condition F (misalignment, 0.4s delay; \textit{M} = 59.72, \textit{SD} = 20.94). 

\begin{figure}[ht]
    \centering
    \includegraphics[width=0.95\columnwidth]{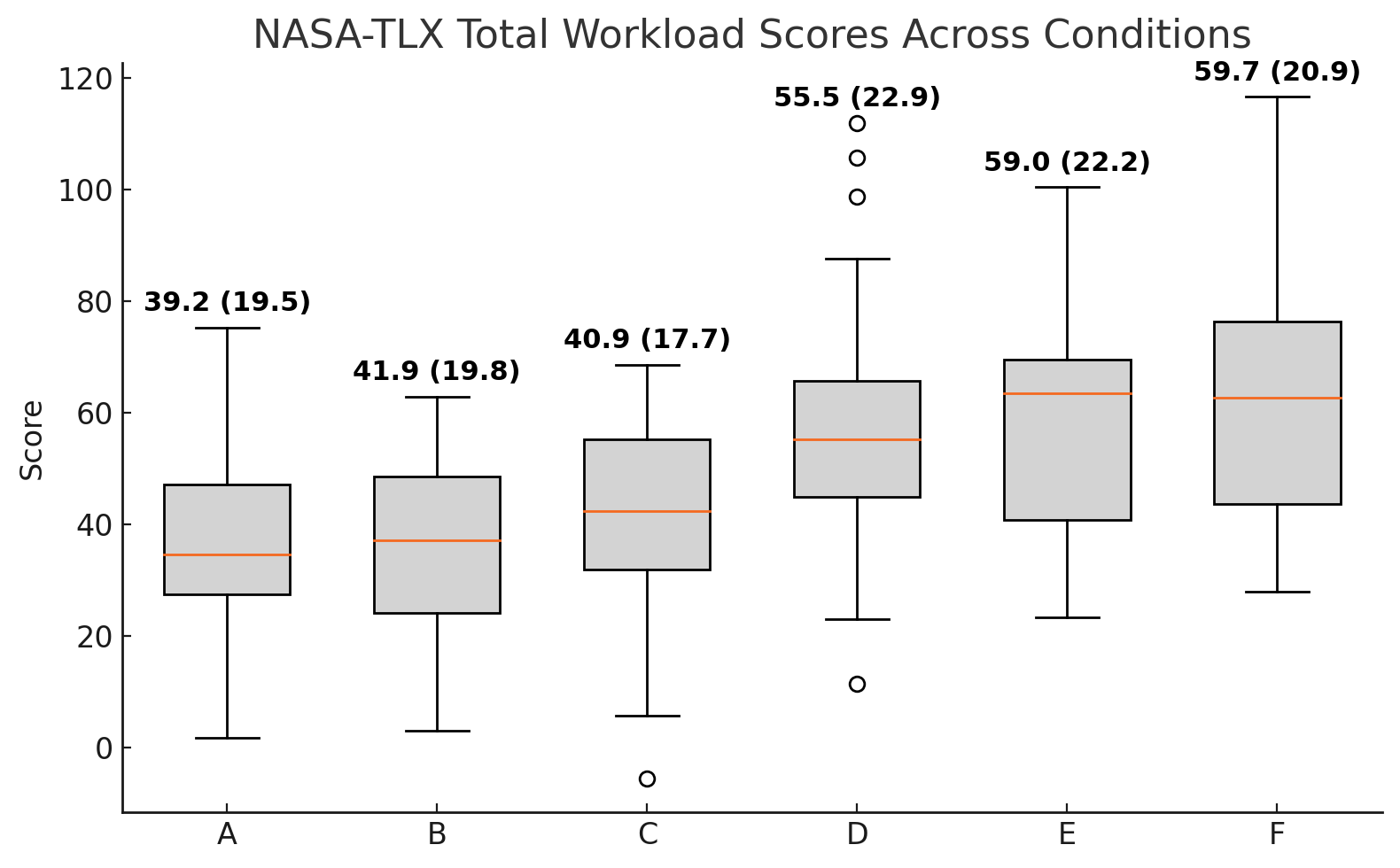}
    \caption{Distribution of NASA-TLX total workload scores across all six experimental conditions (A–F). Each boxplot displays the median (horizontal line), interquartile range (IQR; box), and 1.5 × IQR range (whiskers). Outliers are represented by individual points. Mean values and standard deviations are indicated above each box in the format \textit{M (SD)}. Higher scores indicate higher perceived workload.}
    \label{fig:tlx_total_scores}
\end{figure}

\subsubsection{User Experience Questionnaire (UEQ)}

Table~\ref{tab:ueq_descriptives} displays the descriptive statistics for the Attractiveness, Pragmatic Quality, and Hedonic Quality dimensions of the UEQ across the six experimental conditions. The highest values across all three scales were observed in Condition A (Attractiveness: \textit{M} = 1.59, \textit{SD} = 1.09; Pragmatic Quality: \textit{M} = 1.92, \textit{SD} = 0.28; Hedonic Quality: \textit{M} = 0.62, \textit{SD} = 0.51). According to the UEQ interpretation guidelines, Attractiveness and Pragmatic Quality in this condition were clearly evaluated positively (values $>$ 0.8), while Hedonic Quality remained within the neutral range. Conditions B and C followed a similar pattern, with positive evaluations on Attractiveness and Pragmatic Quality (e.g., Condition B: Attractiveness \textit{M} = 1.39, Pragmatic Quality \textit{M} = 1.88) and neutral values for Hedonic Quality. In contrast, Condition F showed the lowest values for Attractiveness (\textit{M} = 0.30, \textit{SD} = 1.70) and Pragmatic Quality (\textit{M} = 0.11, \textit{SD} = 0.35), both within the neutral range, while Hedonic Quality remained neutral but slightly higher (\textit{M} = 0.53, \textit{SD} = 0.20). Across all conditions, no values fell into the negative evaluation range ($<-0.8$), and Hedonic Quality remained consistently neutral. The descriptive results suggest a gradual decline in perceived Attractiveness and Pragmatic Quality across conditions with increasing misalignment and delay.

\begin{table}[ht]
\centering
\caption{Descriptive statistics for UEQ scales across all experimental conditions. Values represent means with standard deviations.}
\label{tab:ueq_descriptives}
\begin{tabular}{lccc}
\toprule
\textbf{Condition} & \textbf{Attractiveness} & \textbf{Pragmatic Quality} & \textbf{Hedonic Quality} \\
& \textit{M (SD)} & \textit{M (SD)} & \textit{M (SD)} \\
\midrule
\textbf{A} & 1.59 (\textit{1.09}) & 1.92 (\textit{.28}) & .62 (\textit{.51}) \\
\textbf{B} & 1.39 (\textit{1.28}) & 1.88 (\textit{.31}) & .52 (\textit{.46}) \\
\textbf{C} & 1.46 (\textit{1.08}) & 1.76 (\textit{.31}) & .63 (\textit{.44}) \\
\textbf{D} & .67 (\textit{1.59}) & .53 (\textit{.35}) & .74 (\textit{.04}) \\
\textbf{E} & .74 (\textit{1.46}) & .48 (\textit{.40}) & .77 (\textit{.25}) \\
\textbf{F} & .30 (\textit{1.70}) & .11 (\textit{.35}) & .53 (\textit{.20}) \\
\bottomrule
\end{tabular}
\end{table}

\subsection{Repeated-Measures ANOVA}
To examine potential differences between the experimental conditions, repeated-measures analyses of variance (RM-ANOVAs) were conducted for each dependent variable: Attractiveness, Pragmatic Quality, and Hedonic Quality (as measured by the full User Experience Questionnaire, UEQ), as well as perceived workload (as measured by the NASA Task Load Index, NASA-TLX). This analysis approach allows for the detection of within-subject differences across the six experimental conditions and accounts for inter-individual variability. The results provide insight into how user experience and perceived workload may vary as a function of spatial misalignment and system delay. Effect size measures ($\eta^2_{\text{G}}$) are reported to indicate the magnitude of the observed effects and to support interpretation beyond statistical significance.
\vspace{0.5em}

\subsubsection{Repeated-Measures ANOVA: NASA Task Load Index (NASA-TLX)}
A repeated-measures ANOVA was conducted to examine differences in perceived workload (NASA-TLX total score) across the six experimental conditions. The analysis revealed a significant main effect of conditions, $F(5, 155) = 29.09$, $p < .001$, $\eta^2_{\text{G}} = .16$. Mauchly’s test indicated that the assumption of sphericity was violated, $W = .28$, $p < .001$; therefore, Greenhouse–Geisser corrections were applied ($\varepsilon = .70$), which confirmed the significance of the main effect, $F(3.51, 108.25) = 29.09$, $p < .001$.

Bonferroni-corrected pairwise comparisons showed that perceived workload in Conditions D-F (misaligned conditions) was significantly higher than in Conditions A-C (perfect alignment), all $p < .001$. No significant differences were found between Conditions A-C or between Conditions D-F ($p = 1.00$ in all cases).
\vspace{0.5em}

\subsubsection{Repeated-Measures ANOVA: Pragmatic Quality (UEQ)}
The analysis for Pragmatic Quality (UEQ) across the six experimental conditions revealed a significant main effect of conditions, $F(5, 155) = 35.99$, $p < .001$, $\eta^2_{\text{G}} = .36$. Mauchly’s test showed a violation of sphericity, $W = .17$, $p < .001$, so Greenhouse–Geisser corrections were applied ($\varepsilon = .57$), confirming the significance of the effect, $F(2.85, 88.23) = 35.99$, $p < .001$.

Post-hoc pairwise comparisons with Bonferroni adjustments showed that Conditions D, E, and F differed significantly from Conditions A-C ($p < .001$ in all cases). No significant differences were found among Conditions A-C ($p = 1.00$), while comparisons among Conditions D, E, and F also yielded no significant differences ($p = .36$ to $p = 1.00$).
\vspace{0.5em}

\subsubsection{Repeated-Measures ANOVA: Hedonic Quality (UEQ)}
The repeated-measures ANOVA for Hedonic Quality (UEQ) across the six experimental conditions did not reveal a significant main effect of conditions, $F(5, 155) = 0.62$, $p = .68$, $\eta^2_{\text{G}} = .008$. Mauchly’s test indicated a violation of sphericity, $W = .18$, $p < .001$; however, applying the Greenhouse–Geisser correction ($\varepsilon = .65$) did not alter the outcome, $F(3.23, 100.52) = 0.62$, $p = .61$. 

Bonferroni-corrected pairwise comparisons likewise showed no significant differences between any of the conditions ($p = 1.00$ in all cases).
\vspace{0.5em}

\subsubsection{Repeated-Measures ANOVA: Attractiveness (UEQ)}
The analysis for assessing differences in Attractiveness (UEQ) across the six experimental conditions revealed a significant main effect of conditions, $F(5, 155) = 11.12$, $p < .001$, $\eta^2_{\text{G}} = .11$. Mauchly’s test indicated that the assumption of sphericity was violated, $W = .11$, $p < .001$, so Greenhouse–Geisser corrections were applied ($\varepsilon = .55$), which confirmed the significant effect, $F(2.76, 85.88) = 11.12$, $p < .001$.

Bonferroni-corrected pairwise comparisons showed that Attractiveness ratings in Conditions D, E, and F were significantly lower than those in Conditions A-C. Specifically, comparisons such as Condition A vs. D ($p = .010$), A vs. E ($p = .023$), and A vs. F ($p = .002$) all reached statistical significance. No significant differences were found among Conditions A–C ($p = 1.00$), while comparisons between Conditions D-F also yielded non-significant results ($p \geq .43$).
\vspace{0.5em}

Table~\ref{tab:anova_summary} summarizes the repeated-measures ANOVAs for perceived workload and user experience. Significant effects of condition were found for workload, Pragmatic quality, and Attractiveness, while no significant differences emerged for Hedonic quality.

\begin{table}[htbp]
\centering
\caption{Summary of Repeated-Measures ANOVAs for NASA-TLX and UEQ Subscales}
\label{tab:anova_summary}
\begin{tabular}{lccrc}
\toprule
\textbf{Measure} & \textbf{df} & \textbf{F} & \textit{p} & $\eta^2_{\text{G}}$ \\
\midrule
NASA-TLX (Total)            & 5, 155   & 29.09   & $<$ .001   & .160 \\
UEQ – Pragmatic Quality     & 5, 155   & 35.99   & $<$ .001   & .361 \\
UEQ – Hedonic Quality       & 5, 155   & 0.62    & .682     & .008 \\
UEQ – Attractiveness        & 5, 155   & 11.12   & $<$ .001   & .109 \\
\bottomrule
\end{tabular}
\vspace{2mm}
\begin{tablenotes}
\small
\item\textit{Note.} Greenhouse-Geisser correction was applied in all cases due to violation of sphericity. $\eta^2_{\text{G}}$ = generalized eta squared.
\end{tablenotes}
\end{table}
\vspace{0.5em}

Table~\ref{tab:pairwise_summary} summarizes the significant pairwise comparisons between experimental conditions A–F for perceived workload and user experience. Comparisons are Bonferroni-corrected, and only statistically significant differences are reported.

\begin{table}[htbp]
\centering
%\begin{threeparttable}
\caption{Significant Pairwise Comparisons Between Conditions (Bonferroni-Corrected)}
\label{tab:pairwise_summary}
\begin{tabular}{lcr}
\toprule
\textbf{Measure} & \textbf{Comparison} & \textit{p} (adj.) \\
\midrule
NASA-TLX (Total)          & A vs. D & $<$ .001 \\
                          & A vs. E & $<$ .001 \\
                          & A vs. F & $<$ .001 \\
                          & B vs. D & $<$ .001 \\
                          & B vs. E & $<$ .001 \\
                          & B vs. F & $<$ .001 \\
                          & C vs. D & $<$ .001 \\
                          & C vs. E & $<$ .001 \\
                          & C vs. F & $<$ .001 \\
UEQ – Pragmatic Quality   & A vs. D & $<$ .001 \\
                          & A vs. E & $<$ .001 \\
                          & A vs. F & $<$ .001 \\
                          & B vs. D & $<$ .001 \\
                          & B vs. E & $<$ .001 \\
                          & B vs. F & $<$ .001 \\
                          & C vs. D & $<$ .001 \\
                          & C vs. E & $<$ .001 \\
                          & C vs. F & $<$ .001 \\
UEQ – Attractiveness      & A vs. D & .010 \\
                          & A vs. E & .023 \\
                          & A vs. F & .002 \\
                          & B vs. D & .007 \\
                          & B vs. F & .012 \\
                          & C vs. D & .030 \\
                          & C vs. F & $<$ .001 \\
\bottomrule
\end{tabular}
\vspace{2mm}
\begin{tablenotes}
\small
\item\textit{Note.} Only significant comparisons are shown ($p$ $<$ .05, Bonferroni-adjusted).\\A–F refer to the six experimental conditions:\\ A: Perfectly Aligned, 0.0s delay; B: Perfectly Aligned, 0.1s delay; C\textbf{:} Perfectly Aligned, 0.4s delay; D\textbf{:} Misaligned, 0.0s delay; E: Misaligned, 0.1s delay; F: Misaligned, 0.4s delay. 
\end{tablenotes}
%\end{threeparttable}
\end{table}
\vspace{0.5em}

\section{Discussion}

\subsection{Summary of Findings}
Our study demonstrated that spatial misalignment significantly increases cognitive load and degrades the practical aspects of user experience in collaborative AR settings. Participants reported notably higher workload scores (as measured by the NASA-TLX) when virtual objects were misaligned (Conditions D–F) compared to perfectly aligned conditions (Conditions A–C). While varying levels of time delay (0s, 0.1s, 0.4s) within each alignment category did not produce statistically significant differences in workload, the subjective evaluations of Pragmatic quality and Attractiveness on the UEQ were clearly diminished under misaligned conditions. In contrast, Hedonic quality—reflecting aspects of enjoyment and stimulation—remained largely unaffected by either spatial misalignment or time delay.

\subsection{Interpretation of Findings}
The elevated workload in misaligned conditions likely reflects the additional cognitive effort required for participants to reconcile conflicting spatial cues. When virtual objects are not properly aligned, users must continuously recalibrate their mental representations of the task environment, which contributes to increased perceived effort. Similarly, the degradation in Pragmatic quality and Attractiveness ratings for misaligned conditions suggests that accurate spatial positioning is critical for ensuring a smooth, intuitive interaction in AR systems. Interestingly, the stability of Hedonic quality ratings across all conditions implies that the intrinsic enjoyment derived from the task is less sensitive to these technical imperfections. Overall, our findings emphasize that while spatial accuracy is crucial for efficient and satisfying AR interactions, the enjoyment factor may persist even under suboptimal conditions.

\subsection{Limitations and Future Research Directions}
Several limitations warrant consideration. First, although our study involved 32 participants (16 pairs), the relatively modest sample size might have limited our ability to detect more subtle effects, particularly regarding the influence of time delay. Second, the controlled lab environment and the specific task setup may not have been optimal for highlighting the impact of temporal delays; a faster-paced or more dynamic task might better reveal such effects. Additionally, the ecological validity of our findings is constrained by the artificial nature of the experimental conditions, and the scope of measures was limited to quantitative assessments of workload and user experience. Finally, the technical parameters employed—specifically, the degrees of misalignment and the chosen delay intervals—may not fully represent the range of challenges encountered in real-world AR systems.

Future research should overcome these limitations by employing larger, more diverse samples and by exploring alternative experimental paradigms that more closely mimic real-world interactions. Expanding the range of technical parameters (e.g., additional time delay intervals, different spatial misalignment patterns, randomized jitter, etc.) and incorporating additional performance metrics (e.g., error rates, reaction times, or physiological measures, etc.) could provide a more comprehensive understanding of how spatial misalignment and time delay affect collaborative AR experiences. Finally, conducting studies in ecologically valid environments would further enhance the generalizability of these findings and help to establish guidelines for the design and development of more robust AR systems.

\section{Conclusion}
This study investigated the impact of spatial misalignment and time delay on collaborative AR interactions. Our findings reveal that spatial misalignment significantly increases cognitive load and degrades practical aspects of the user experience, as evidenced by higher NASA-TLX scores and lower ratings in Pragmatic quality and Attractiveness. In contrast, variations in time delay within each alignment category had a less pronounced effect and the Hedonic quality remained stable under all conditions.

These results underscore the importance of precise virtual object alignment in the design of collaborative AR systems. Enhancing spatial accuracy can improve usability and efficiency, thereby reducing the cognitive burden on users. Although task-related enjoyment may persist despite technical imperfections, ensuring clear and reliable spatial representations is critical for effective communication and task performance.

Future work should address the limitations of the current study by employing larger and more diverse samples, exploring a broader range of technical parameters, and incorporating additional performance metrics in ecologically valid settings. Such research will help further elucidate the complex interplay between technical factors and user experience in AR, ultimately guiding the development of more robust and user-friendly systems.

In summary, our study contributes to the understanding of how spatial misalignment and time delay affect collaborative AR environments and provides a foundation for future research aimed at optimizing these immersive technologies.

\section*{Acknowledgment}
The authors used ChatGPT4.0 and Grammarly for grammar and spell-checking. After using these tool(s)/service(s), the authors reviewed and edited the content as needed and take full responsibility for the publication’s content.  

\bibliographystyle{IEEEtran}
\bibliography{conference_101719.bib}
%\begin{thebibliography}{00}
%\bibitem{b1} G. Eason, B. Noble, and I. N. Sneddon, ``On certain integrals of Lipschitz-Hankel type involving products of Bessel functions,'' Phil. Trans. Roy. Soc. London, vol. A247, pp. 529--551, April 1955.
%\bibitem{b2} J. Clerk Maxwell, A Treatise on Electricity and Magnetism, 3rd ed., vol. 2. Oxford: Clarendon, 1892, pp.68--73.
%\bibitem{b3} I. S. Jacobs and C. P. Bean, ``Fine particles, thin films and exchange anisotropy,'' in Magnetism, vol. III, G. T. Rado and H. Suhl, Eds. New York: Academic, 1963, pp. 271--350.
%\bibitem{b4} K. Elissa, ``Title of paper if known,'' unpublished.
%\bibitem{b5} R. Nicole, ``Title of paper with only first word capitalized,'' J. Name Stand. Abbrev., in press.
%\bibitem{b6} Y. Yorozu, M. Hirano, K. Oka, and Y. Tagawa, ``Electron spectroscopy studies on magneto-optical media and plastic substrate interface,'' IEEE Transl. J. Magn. Japan, vol. 2, pp. 740--741, August 1987 [Digests 9th Annual Conf. Magnetics Japan, p. 301, 1982].
%\bibitem{b7} M. Young, The Technical Writer's Handbook. Mill Valley, CA: University Science, 1989.
%\end{thebibliography}
%\vspace{12pt}
%\color{red}

\end{document}